\documentstyle[preprint,aps,floats,epsf]{revtex}

\begin{document}

\tightenlines %% LANL requests single line spacing to save paper

%------------------------------------
% New Commands	
%------------------------------------
\newcommand{\notE}{\ \hbox{{$E$}\kern-.60em\hbox{/}}}
\newcommand{\notp}{\ \hbox{{$p$}\kern-.43em\hbox{/}}}

%=======================================================================
%   TITLE PAGE
%=======================================================================

\includegraphics{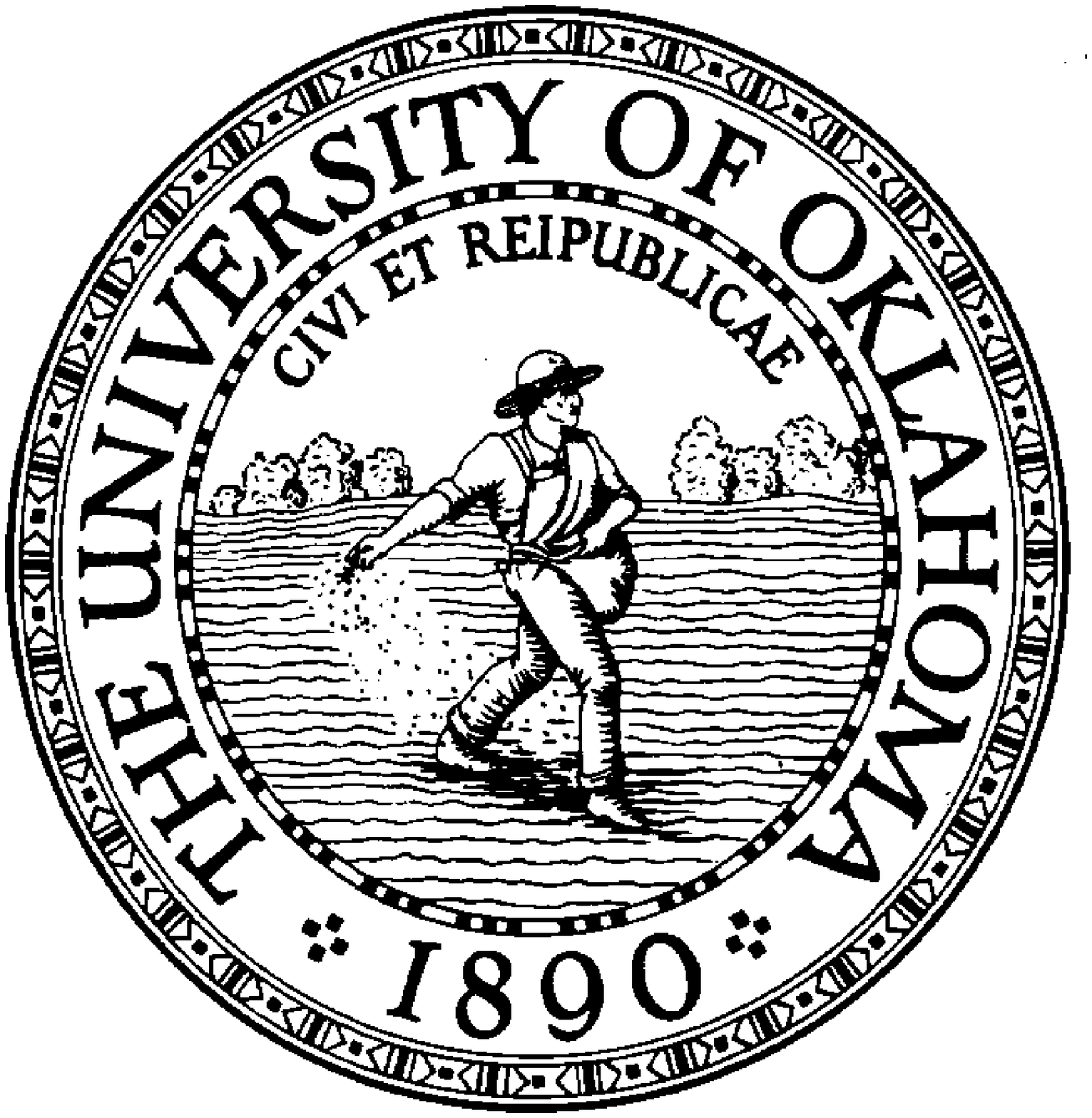}

%-----------------------------------
%   Preprint Number
%-----------------------------------
\preprint{\font\fortssbx=cmssbx10 scaled \magstep2
\hbox to \hsize{
\hskip1.2in %\raise.1in
\hbox{\fortssbx The University of Oklahoma}
\hskip0.8in $\vcenter{
                      \hbox{\bf BNL-HET-04/01}
                      \hbox{\bf OKHEP-04-01}
                      \hbox{\bf UTHEP-04-01}
                      \hbox{\bf hep-ph/0402172}
                      \hbox{February 2004}}$ }
}
 
%-----------------------------------
%   Title
%-----------------------------------
\title{\vspace{.5in}
Discovering the Higgs Bosons of Minimal Supersymmetry \\ 
with Muons and a Bottom Quark}
 
%-----------------------------------
%   Authors
%-----------------------------------
\author{
Sally Dawson$^a$, Duane Dicus$^b$, Chung Kao$^c$ and Rahul Malhotra$^b$}

%-----------------------------------
%   Address
%-----------------------------------
\address{
$^a$Department of Physics, Brookhaven National Laboratory, 
Upton, NY 11973, USA \\
$^b$Center for Particle Physics, University of Texas, 
Austin, TX 78712, USA \\
$^c$Department of Physics and Astronomy, University of Oklahoma, 
Norman, OK 73019, USA}

\maketitle

\bigskip

%-----------------------------------
%   Abstract
%-----------------------------------
\begin{abstract}

We investigate the prospects for the discovery at the CERN Large Hadron 
Collider of a neutral Higgs boson produced with one bottom quark followed 
by Higgs decay into a muon pair.  
We work within the framework of the minimal supersymmetric model.
The dominant physics background from the production of $b\mu^+\mu^-$, 
$j\mu^+\mu^-, j = g, u, d, s, c$, and $b\bar{b}W^+W^-$ is calculated 
with realistic acceptance cuts.
Promising results are found for the CP-odd pseudoscalar ($A^0$) 
and the heavier CP-even scalar ($H^0$) Higgs bosons with masses up to 600 GeV.
This discovery channel with one energetic bottom quark greatly improves 
the discovery potential of the LHC beyond the inclusive channel 
$pp \to \phi^0 \to \mu^+\mu^- +X$.

\end{abstract}

%

%=======================================================================
%   BEGIN MAIN TEXT
%=======================================================================
\newpage

%-----------------------------------------------------------------------
 {\bf I. Introduction}
%-----------------------------------------------------------------------
%{\sl Introduction.}--
\newline
In the minimal supersymmetric standard model (MSSM) \cite{MSSM}, 
the Higgs sector has Yukawa interactions with two doublets 
$\phi_1$ and $\phi_2$ that couple to fermions with weak isospin 
$-1/2$ and $+1/2$ respectively \cite{Guide}. 
After spontaneous symmetry breaking, there remain five physical Higgs bosons:
a pair of singly charged Higgs bosons $H^{\pm}$,
two neutral CP-even scalars $H^0$ (heavier) and $h^0$ (lighter),
and a neutral CP-odd pseudoscalar $A^0$.
The Higgs potential is constrained by supersymmetry 
such that all tree-level Higgs boson masses and couplings 
are determined by just two independent parameters,  
commonly chosen to be the mass of the CP-odd pseudoscalar ($m_A$) 
and the ratio of vacuum expectation values of neutral Higgs fields 
($\tan\beta \equiv v_2/v_1$). 

At the CERN Large Hadron Collider (LHC), 
gluon fusion ($gg \to \phi^0, \phi^0 = h^0, H^0$, or $A^0$) is the major 
source of neutral Higgs bosons in the MSSM for $\tan\beta$ less than about 5.
If $\tan\beta$ is larger than 7, neutral Higgs bosons are dominantly 
produced from bottom quark fusion $b\bar{b} \to \phi^0$ 
\cite{Dicus1,Dicus2,Balazs,Maltoni,Harlander}. 
Since the Yukawa couplings of $\phi^0 b\bar{b}$ are enhanced by $1/\cos\beta$,
the production rate of neutral Higgs bosons, especially the $A^0$ 
or the $H^0$, is enhanced at large $\tan\beta$.

%-----------------------------------
% tau tau and bb
%-----------------------------------
For large $\tan\beta$, 
the $\tau\bar{\tau}$ decay mode \cite{Kunszt,Richter-Was}
is a promising discovery channel for the $A^0$ and the $H^0$ in the MSSM.
%-----------------------------------
% mu mu
%-----------------------------------
The LHC discovery potential of the inclusive muon pair channel 
for neutral Higgs bosons of minimal supersymmetry 
was demonstrated by Kao and Stepanov \cite{Nikita,CMS},  
and was later confirmed by the ATLAS collaboration \cite{ATLAS}.  
In the minimal supergravity unified model \cite{mSUGRA}, 
the significance of $pp \to \phi^0 \to \mu^+\mu^-+X$ is 
greatly improved by a large $\tan\beta$ \cite{Vernon} because  
the large $b\bar{b}\phi^0$ couplings make $m_A$ and $m_H$ small through 
the evolution of renormalization group equations \cite{Baer}.

For a Higgs boson produced along with one bottom quark at high transverse 
momentum ($p_T$), the leading-order subprocess is 
$bg \to b \phi^0$ \cite{Choudhury,Huang,Scott,Cao}. 
If two high $p_T$ bottom quarks are required in association with 
a Higgs boson, the leading order subprocess should be 
$gg \to b\bar{b}\phi^0$ \cite{Dicus1,hbbmm,Plumper,Dittmaier,Dawson}. 

%-----------------------------------
% Summary
%-----------------------------------
Recently, it has been suggested that the search at the LHC for a Higgs boson 
produced along with a single bottom quark with large $p_T$ should be 
more promising than the production of a Higgs boson associated with 
two high $p_T$ bottom quarks \cite{Scott}.
In this letter, we present the prospects of discovering the MSSM neutral 
Higgs bosons produced with a bottom quark via Higgs decays into muon pairs. 
We calculate the Higgs signal and the dominant Standard Model (SM) background 
of $b \mu^+\mu^-, b = b$ or $\bar{b}$, with realistic cuts and compare 
its discovery potential with that of the inclusive final state of 
$pp \to \phi^0 \to \mu^+\mu^- +X, \phi^0 = H^0, h^0, A^0$. 
This discovery channel with one energetic bottom quark greatly improves  
the discovery potential beyond the reach of 
not only the associated mode with two bottom quarks 
$pp \to b\bar{b}\phi^0 \to b\bar{b}\mu^+\mu^- +X$ \cite{hbbmm} 
but also the inclusive channel $pp \to \phi^0 \to \mu^+\mu^- +X$. 

%------------------------------------------------------------------------------
 {\bf II. The production cross sections and branching fractions}
%------------------------------------------------------------------------------
\newline
%{\sl The production cross sections and branching fractions.}--
We calculate the cross section at the LHC for $pp \to b \phi^0 +X$ 
($\phi^0 = H^0, h^0, A^0$) via $bg \to b \phi^0$ 
with the parton distribution functions of CTEQ6L1 \cite{CTEQ6}. 
The factorization scale is chosen to be $M_H/4$ \cite{Maltoni,Plehn}.
In this letter, unless explicitly specified,
$b$ represent a bottom quark ($b$) or an anti-bottom quark ($\bar{b}$). 
The bottom quark mass in the $\phi^0 b\bar{b}$ Yukawa coupling 
is chosen to be the NLO running mass $m_b(\mu_R)$ \cite{bmass}, 
which is calculated with $m_b({\rm pole}) = 4.7$ GeV and the NLO evolution 
of the strong coupling \cite{alphas}. 
We have also taken the renormalization scale to be $M_H/4$.  
This choice of scale effectively reproduces the effects of next-to-leading 
order (NLO) \cite{Scott}. Therefore, we take the $K$ factor to be one 
for the Higgs signal. 

The cross section for $pp \to b \phi^0 \to b \mu^+\mu^- +X$ can be thought of 
as the Higgs production cross section $\sigma(pp \to b \phi^0 +X)$ 
multiplied by the branching fraction of the Higgs decay into muon pairs
$B(\phi^0 \to \mu^+\mu^-)$.
(In reality we integrate over a $\mu^+ \mu^-$ invariant mass bin centered
on $m_{\phi}$.)
When the $b\bar{b}$ mode dominates Higgs decays, 
the branching fraction of $\phi^0 \to \mu^+\mu^-$ is about 
$m_\mu^2/3 m_b^2(m_\phi)$
where $m_b(m_\phi)$ is the running mass at the scale $m_\phi$.  
This results in a branching fraction for $A^0 \to \mu^+\mu^-$ 
of approximately $3 \times 10^{-4}$ for $m_A = 100$ GeV.
For $\tan\beta \agt 10$ and $m_A \agt 125$ GeV, the cross section of $bA^0$ 
or that of $bH^0$ is  enhanced by approximately $\tan^2\beta$; 
the muon branching fraction is sustained by the large decay 
width of the Higgs  into bottom quarks.

%------------------------------------------------------------------------------
 {\bf III. The Physics Background}
%------------------------------------------------------------------------------
%{\sl The Physics Background.}--
\newline
The dominant physics backgrounds to the final state of $b \mu^+\mu^-$ 
come from $bg \to b \mu^+\mu^-$ ($b\mu\mu$) 
as well as $gg \to b\bar{b}W^+W^-$ and $q\bar{q} \to b\bar{b}W^+W^-$ ($bbWW$) 
followed by the decays of $W^\pm \to \mu^\pm \nu_\mu$.
In our analysis, we actually evaluated 
$gg \to b\mu^+\nu_\mu \bar{b}\mu^-\bar{\nu}_\mu$ and 
$q\bar{q} \to b\mu^+\nu_\mu \bar{b}\mu^-\bar{\nu}_\mu$ whose 
dominant contribution is from $pp \to t\bar{t} \to bW^+\bar{b}W^- +X$.
In addition, we have included the background from 
$bg \to b\mu^+\nu \mu^-\bar{\nu}$ and
$\bar{b}g \to \bar{b}\mu^-\bar{\nu} \mu^+\nu$ 
which has major contributions from 
$bg \to tW^-$ and $\bar{b}g \to \bar{t}W^+$ ($tW$). 
The cross section of the $tW$ background is approximately $1/10$ that of 
$bbWW$.
The muons from $b$ decays can be removed effectively with isolation cuts 
\cite{Nikita}.
We have applied a K factor of 1.3 for the $b\mu\mu$ background 
\cite{Campbell}, a K factor of 2 for $bbWW$ \cite{Bonciani,Nason}, 
and a K factor of 1.5 for $tW$ \cite{Zhu}. 
The Feynman diagrams for $bg \to b \mu^+\mu^-$ are shown in Fig. 1.
We have also considered backgrounds from 
$pp \to j \mu^+\mu^- +X, j = g, q$ or $\bar{q}$ with $q = u, d, s, c$, 
where a jet is mistagged as a $b$;  
we use a K factor of 1.3 for these processes.

In every event, each of the two isolated muons is required to have
$p_T(\mu) > 20$ GeV and $|\eta(\mu)| < 2.5$. 
For an integrated luminosity ($L$) of 30 fb$^{-1}$, 
we require $p_T(b,j) > 15$ GeV and $|\eta(b,j)| < 2.5$. 
The $b$-tagging efficiency ($\epsilon_b$) is taken to be $60\%$, 
the probability that a $c$-jet is mistagged as a $b$-jet ($\epsilon_c$)
is $10\%$ and 
the probability that any other jet is mistagged as a $b$-jet ($\epsilon_j$)
is taken to be $1\%$.

For a higher integrated luminosity of 300 fb$^{-1}$, 
we require the same acceptance cuts as  for $L =$ 30 fb$^{-1}$ 
except for $p_T(b,j) > 30$ GeV and $\epsilon_b =50\%$.
In addition, to reduce the background from $bbWW$ and $tW$
which contains neutrinos, we require that 
the missing transverse energy ($\notE_T$) in each event 
should be less than 20 GeV for $L = 30$ fb$^{-1}$ 
and less than 40 GeV for $L = 300$ fb$^{-1}$. 

We have employed the programs MADGRAPH \cite{Madgraph}
and HELAS \cite{Helas} to evaluate
the background cross sections of 
$pp \to b\mu^+\mu^- +X, j\mu^+\mu^- +X$ and $bbW^+W^- +X$.
The background from $bbW^+W^-$ are treated with special care for $b-$tagging. 
If there is only one $b$ passing the cuts, the cross section is multiplied 
with $\epsilon_b$. For the events with two $b$'s passing the cuts, 
we multiply the cross section with $2\epsilon_b -\epsilon_b^2$. 

We have compared the prospects of detecting this Higgs signal 
with one high $p_T$ bottom quark ($pp \to b\phi^0 \to b\mu^+\mu^- +X$)  
with that of the inclusive channel $pp \to \phi^0 \to \mu^+\mu^- +X$ 
and the associated discovery mode with two high $p_T$ bottom quarks 
$pp \to b\bar{b} \phi^0 \to b\bar{b} \mu^+\mu^- +X$.
The associated Higgs signal $b\bar{b} \phi^0 \to b\bar{b}\mu^+\mu^-$ 
has major physics background from 
$pp \to b\bar{b} W^+ W^- +X$, 
$pp \to b\bar{b} \mu^+\mu^- +X$, and  
$pp \to jj \mu^+\mu^- +X$ \cite{hbbmm}.  
The dominant physics background to the inclusive final state of $\mu^+\mu^-$ 
comes from the Drell-Yan process $q\bar{q} \to Z,\gamma \to \mu^+\mu^-$ 
\cite{Nikita}. 

%------------------------
% NEW PAGE
%------------------------
\newpage

%------------------------------------------------------------------------------
 {\bf IV. The Discovery Potential at the LHC}
%------------------------------------------------------------------------------
%{\sl The Discovery Potential at the LHC.}--
\newline
To study the discovery potential of 
$pp \to b \phi^0 +X \to b \mu^+\mu^- +X$ at the LHC, 
we calculate the background from the SM processes of 
$pp \to b \mu^+\mu^- +X$
in the mass window of
$m_\phi \pm \Delta M_{\mu^+\mu^-}$ where 
$\Delta M_{\mu^+\mu^-} \equiv 
1.64 [ (\Gamma_\phi/2.36)^2 +\sigma_m^2 ]^{1/2}$ \cite{ATLAS}.
$\Gamma_\phi$ is the total width of the Higgs boson,  
and $\sigma_m$ is the muon mass resolution which
we take to be $2\%$ of the Higgs boson mass \cite{ATLAS}.
The CMS mass resolution will be better than $2\%$ of $m_\phi$ for 
$m_\phi \alt$ 500 GeV \cite{Nikita,CMS}. 
Therefore, the observability for the muon pair discovery channel 
at the CMS detector will be better than what is shown in this letter.

In Figure 2 we show the cross section of muon pairs from Higgs decays 
along with a bottom quark, $\sigma(pp \to b A^0 +X \to b \mu^+\mu^- +X)$, 
for $\tan\beta = 10$ and 50, with a common mass for scalar quarks, scalar 
leptons and the gluino $m_{\tilde{q}} = m_{\tilde{g}} = \mu = 1$ TeV.
We also present the background cross sections in the mass window of 
$m_A \pm \Delta M_{\mu^+\mu^-}$ for the SM processes 
$pp \to b \mu^+\mu^- +X$, $pp \to j \mu^+\mu^- +X$, 
and $pp \to b\bar{b} W^+ W^- +X$. 
The cuts, tagging efficiencies, and  K factors discussed above are included. 
There are a couple of things to note from this figure. 
\begin{itemize}
\item[(a)] For an integrated luminosity of 30 fb$^{-1}$, 
the cross section of the Higgs signal with $\tan\beta \sim 50$ 
can be much larger than that of the physics background after acceptance cuts.  
The SM subprocesses  $gg \to b \mu^+\mu^-$ and 
$q\bar{q} \to b \mu^+\mu^-$ make the major contributions 
to the physics background for $M_{\mu^+\mu^-} \alt 180$ GeV,
but $gg \to b\bar{b}W^+W^-$ and $q\bar{q} \to b\bar{b}W^+W^-$ 
become the dominant background for higher muon pair invariant mass. 
\item[(b)] At the higher luminosity of 300 fb$^{-1}$,  
$gg \to b\bar{b}W^+W^-$ and $q\bar{q} \to b\bar{b}W^+W^-$ 
make up the dominant background for $M_{\mu^+\mu^-} \agt 120$ GeV. 
The higher $p_T$ cut on the $b-$quark reduces the Higgs signal, 
while the larger allowed missing $E_T$ make the $bbWW$ background 
greater than the Higgs signal with $\tan\beta \alt 50$.
\end{itemize}

We define the signal to be observable 
if the lower limit on the signal plus background is larger than 
the corresponding upper limit on the background \cite{HGG,Brown}, namely,
\begin{eqnarray}
L (\sigma_s+\sigma_b) - N\sqrt{ L(\sigma_s+\sigma_b) } > 
L \sigma_b +N \sqrt{ L\sigma_b }
\end{eqnarray}
which corresponds to
\begin{eqnarray}
\sigma_s > \frac{N^2}{L} \left[ 1+2\sqrt{L\sigma_b}/N \right]
\end{eqnarray}
Here $L$ is the integrated luminosity, 
$\sigma_s$ is the cross section of the Higgs signal, 
and $\sigma_b$ is the background cross section.  
Both cross sections are taken to be 
within a bin of width $\pm\Delta M_{\mu^+\mu^-}$ centered at $m_\phi$. 
In this convention, $N = 2.5$  corresponds to a 5$\sigma$ signal.
We take the integrated luminosity $L$ to be 30 fb$^{-1}$ 
and 300 fb$^{-1}$ \cite{ATLAS}. 

For $\tan\beta \agt 10$, 
$m_A$ and $m_H$ are almost degenerate when $m_A \agt$ 125 GeV, 
while $m_A$ and $m_h$ are very close 
to each other for $m_A \alt$ 125 GeV in the MSSM. 
Therefore, when computing the discovery reach, we add the cross sections 
of the $A^0$ and the $h^0$ for $m_A < 125$ GeV 
and those of the $A^0$ and the $H^0$ for $m_A \ge 125$ GeV 
\cite{Nikita,CMS,ATLAS}.

Figure 3 shows the 5$\sigma$ discovery contours for the MSSM Higgs bosons 
where the discovery region is the part of the parameter space above the 
contour. We have chosen 
$M_{\rm SUSY} = m_{\tilde{q}} = m_{\tilde{g}} = m_{\tilde{\ell}} 
= \mu = 1$ TeV.
If $M_{\rm SUSY}$ is smaller, the discovery region of $A^0,H^0 \to \mu^+\mu^-$ 
will be slightly reduced for $m_A \agt 250$ GeV,
because  the Higgs bosons can decay into SUSY particles \cite{HZ2Z2} 
and the branching fraction of $\phi^0 \to \mu^+\mu^-$ is suppressed.
For $m_A \alt 125$ GeV, the discovery region of $H^0 \to \mu^+\mu^-$ 
is slightly enlarged for a smaller $M_{\rm SUSY}$, 
but the observable region of $h^0 \to \mu^+\mu^-$ is slightly reduced 
because the lighter top squarks make the $H^0$ and the $h^0$ lighter; 
also the $H^0 b\bar{b}$ coupling is enhanced 
while the $h^0 b\bar{b}$ coupling is reduced \cite{Nikita,Vernon}.

%-----------------------------------------------------------------------
 {\bf V. Conclusions}
%-----------------------------------------------------------------------
%{\sl Conclusions.}--
\newline
The muon pair decay mode is a promising channel for the discovery of 
the neutral Higgs bosons in the minimal supersymmetric model at the LHC. 
The $A^0$ and the $H^0$ should be observable in a large region 
of parameter space with $\tan\beta \agt 10$.
In particular, Fig. 3 shows that the associated final state of 
$b\phi^0 \to b\mu^+\mu^-$ 
could discover the $A^0$ and the $H^0$ at the LHC 
with an integrated luminosity of 30 fb$^{-1}$ if $m_A \alt 600$ GeV.
At a higher luminosity of 300 fb$^{-1}$, the discovery region in $m_A$ 
is expanded up to $m_A \alt 800$ GeV for $\tan\beta \sim 50$.
This discovery channel with one energetic bottom quark extends the discovery 
potential of the LHC beyond the inclusive channel 
$pp \to \phi^0 \to \mu^+\mu^- +X$.

The excellent muon mass resolution of the CMS and the ATLAS detectors 
will be important for Higgs searches at the LHC.
For large $\tan\beta$, the muon pair discovery mode might 
be the only channel at the LHC that allows precise reconstruction
of the $A^0$ and the $H^0$ masses. 
The discovery of the associated final states of 
$b\phi^0 \to b\mu^+\mu^-$ and $b\bar{b}\phi^0 \to b\bar{b}\mu^+\mu^-$ 
will provide information about the Yukawa couplings of 
$b\bar{b}\phi^0$ and an opportunity to measure $\tan\beta$. 
The discovery of both $\phi^0 \to \tau\bar{\tau}$ and $\phi^0 \to \mu^+\mu^-$ 
will allow us to understand the Higgs Yukawa couplings with the leptons.

%------------------------------------------------------------------------------
  {\bf Acknowledgments}
%------------------------------------------------------------------------------
%{\sl Acknowledgments.}--
\newline
We are grateful to Tim Stelzer for providing a FORTRAN code of MADGRAPH 
that can calculate processes with more than five particles in the final state, 
and to Scott Willenbrock for continuing encouragement.
R.M. is grateful to John Campbell for his help in using the program MCFM 
for NLO calculations of the $bH$ process. 
S.D. would like to thank Laura Reina and Doreen Wackeroth for discussions. 
C.K. would like to thank John Ellis and CERN Theoretical Physics Division, 
where part of the research was completed, for their hospitality.
This research was supported in part by the U.S. Department of Energy
under Grants 
No.~DE-AC02-98CH10886,
No.~DE-FG03-98ER41066, 
No.~DE-FG02-03ER46040,
and
No.~DE-FG03-93ER40757.
 
%-----------------------------------------------------------------------
%   REFERENCES
%-----------------------------------------------------------------------
%

%-----------------------------------------------------------------------
%   FIGURES
%-----------------------------------------------------------------------

%------------------------------------------------
% FIG. 1
%------------------------------------------------

\begin{figure}
\centering\leavevmode
\epsfxsize=6in\epsffile{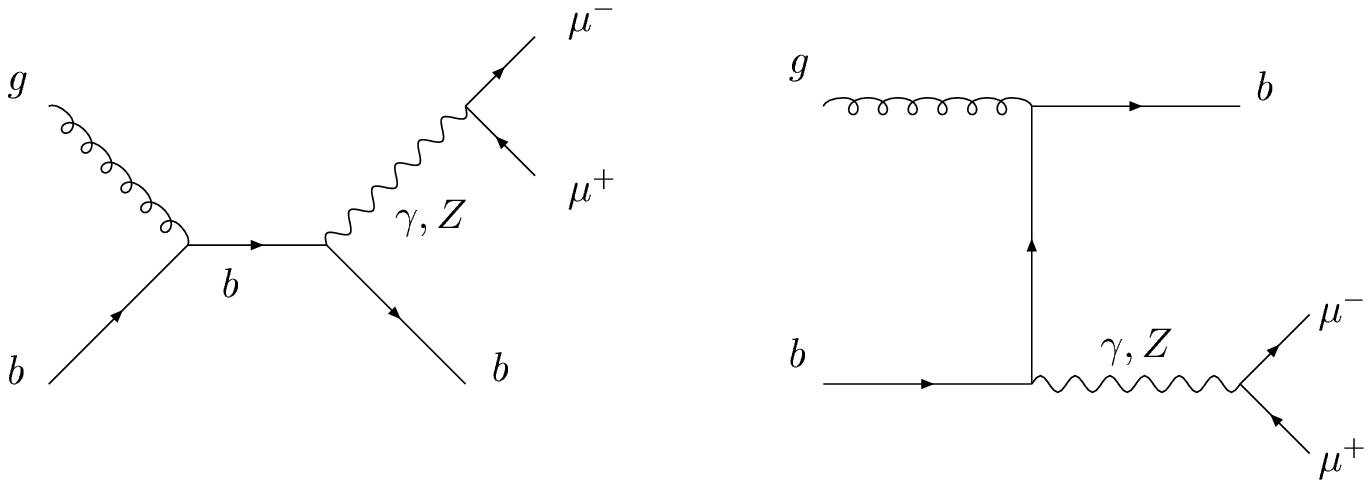}

\caption[]{
Feynman diagrams for the background from $bg \to b \mu^+\mu^-$.
\label{fig:Diagrams}
}\end{figure}
%

%------------------------------------------------
% FIG. 2
%------------------------------------------------

\begin{figure}
\centering\leavevmode
\epsfxsize=6in\epsffile{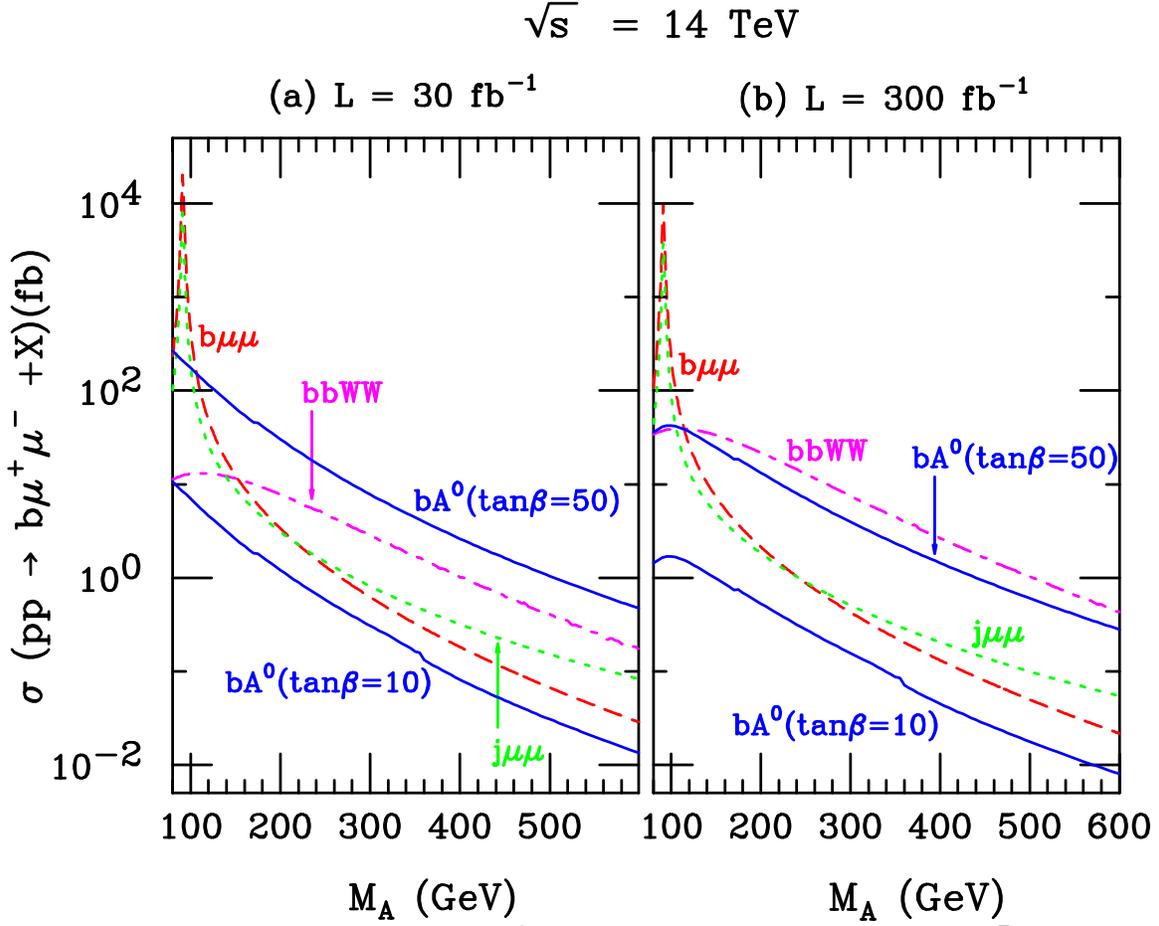}

\caption[]{
The cross section of 
$pp \to bA^0 +X \to b\mu^+\mu^- +X$, $b = b \;\; {\rm or} \;\; \bar{b}$, 
at the LHC, as a function of $m_A$, for
$m_{\tilde{q}} = m_{\tilde{g}} = \mu = 1$ TeV and $\tan\beta = 10$ or 50.
Also shown are the background cross sections in the mass window of 
$m_A \pm \Delta M_{\mu^+\mu^-}$ as discussed in the text for the SM processes 
$pp \to b\mu^+\mu^- +X, b = b \;\; {\rm or } \;\; \bar{b}$, (dashed),
$pp \to j\mu^+\mu^- +X, j = g, u, d, s, c$ (dotted), and
$pp \to b\bar{b}W^+W^- +X$ (dot-dashed).
We have applied K factors, acceptance cuts, and efficiencies of b tagging 
and mistagging. 
\label{fig:sigma}
}\end{figure}

%------------------------------------------------
% FIG. 3
%------------------------------------------------

\begin{figure}
\centering\leavevmode
\epsfxsize=6in\epsffile{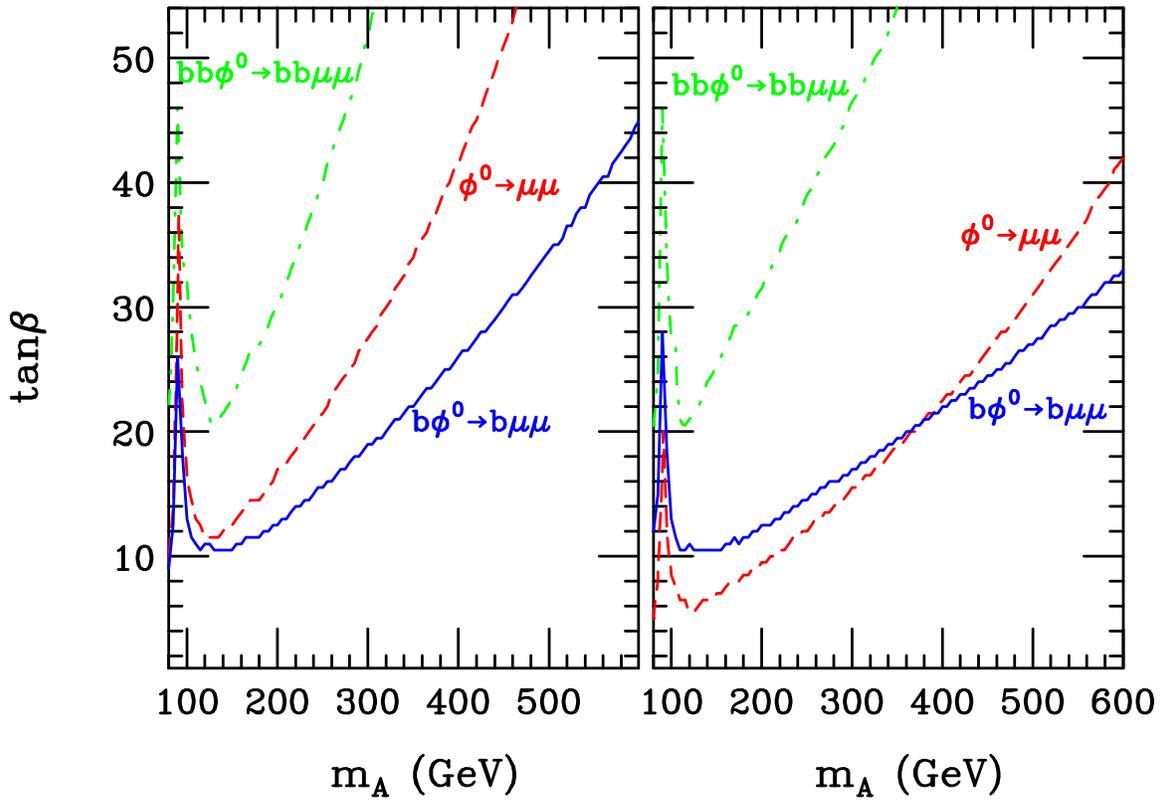}

\caption[]{
The $5\sigma$ discovery contours at the LHC 
for an integrated luminosity ($L$) of 30 fb$^{-1}$ and 300 fb$^{-1}$ 
in the $m_A$ versus $\tan\beta$ plane.  
The signal includes $\phi^0 = A^0$ and $h^0$ for $m_A < 125$ GeV, 
and $\phi^0 = A^0$ and $H^0$ for $m_A \ge 125$ GeV.
The discovery region is the part of the parameter space above the contours.
\label{fig:contour}
}\end{figure}
%

%-----------------------------------------------------------------------
%   END DOCUMENT
%-----------------------------------------------------------------------
\end{document}